# The Superconducting Transition Temperatures of $Fe_{1+x}Se_{1-y}$, $Fe_{1+x}Se_{1-y}Te_y$ and $(K/Rb/Cs)_zFe_{2-x}Se_2$

Dale R. Harshman[1,2,] and Anthony T. Fiory[3]

[1] *Physikon Research Corporation, Lynden, WA 98264 USA*

[2] *Department of Physics, University of Notre Dame, Notre Dame, IN 46556 USA*

[3] *Department of Physics, New Jersey Institute of Technology, Newark, NJ 07102 USA*

Email: drh@physikon.net

**Abstract**

In a recent contribution to this journal, it was shown that the transition temperatures of optimal high-$T_C$ compounds obey the algebraic relation, $T_{C0} = k_B^{-1}\beta/\ell\zeta$, where $\ell$ is related to the mean spacing between interacting charges in the layers, $\zeta$ is the distance between interacting electronic layers, $\beta$ is a universal constant and $k_B$ is Boltzmann's constant. The equation was derived assuming pairing based on interlayer Coulomb interactions between physically separated charges. This theory was initially validated for 31 compounds from five different high-$T_C$ families (within an accuracy of ±1.37 K). Herein we report the addition of $Fe_{1+x}Se_{1-y}$ and $Fe_{1+x}Se_{1-y}Te_y$ (both optimized under pressure) and $A_zFe_{2-x}Se_2$ (for A = K, Rb, or Cs) to the growing list of Coulomb-mediated superconducting compounds in which $T_{C0}$ is determined by the above equation. Doping in these materials is accomplished through the introduction of excess Fe and/or Se deficiency, or a combination of alkali metal and Fe vacancies. Consequently, a very small number of vacancies or interstitials can induce a superconducting state with a substantial transition temperature. The confirmation of the above equation for these Se-based Fe chalcogenides increases to six the number of superconducting families for which the transition temperature can be accurately predicted.

## 1. Introduction

Superconductivity in the binary Fe chalcogenide, $Fe_{1+x}Se_{1-y}$ (denoted 11), was first reported in 2008 [1] [2].[4] Initial measurements indicated a transition temperature $T_C \sim 8$ K, which was raised to 37 K under hydrostatic pressure [3] [4]. Interest in superconductivity of Te-substituted ternary compounds $Fe_{1+x}Se_{1-y}Te_y$ derives from the non-superconducting end-point FeTe and distinctive structure and $T_C$ behavior (relative to FeSe) associated with the larger Te ion (relative to Se) [5] [6]. Ternary compounds of the form $A_zFe_{2-y}Se_2$ (122), with alkali metals intercalated between the Se-2Fe-Se structures soon followed [7], exhibiting $T_C$ values of order 32 K at ambient pressure [8,9]. The intercalated structures exhibit a high resistivity at $T_C$, indicative of inhomogeneity arising, at least in part, from Fe and alkali metal vacancies.

The binary compound $Fe_{1+x}Se_{1-y}$, in which optimum superconductivity occurs at a small Fe excess or Se deficiency, is justifiably a "high-$T_C$" compound, since application of conventional phonon mediation theory was shown not to account for the dramatic increase in $T_C$ observed under applied pressure [10].

[4] The authors of [1] originally identified the α-phase as the superconducting Fe-Se phase, but later corrected it to the β-phase.

**Table 1.** Analysis data for 36 optimal high-$T_C$ superconducting compounds (the stoichiometry and hydrostatically applied pressure values correspond to optimal transition temperature $T_{C0}$). Listed are the measured $T_{C0}$, the distance between interacting layers $\zeta$, the calculated spacing between interacting charges within layers $\ell$, the structures of types I and II reservoirs per formula unit ($O_x$ denotes partial filling), and the theoretical $T_{C0}$. Structurally- and doping-related compounds are grouped accordingly.

| Superconducting compound | Meas. $T_{C0}$ (K) | $\zeta$ (Å) | $\ell$ (Å) | Type I | Type II | Calc. $T_{C0}$ (K) |
|---|---|---|---|---|---|---|
| $YBa_2Cu_3O_{6.92}$ | 93.7 | 2.2677 | 5.7085 | BaO-CuO-BaO | $CuO_2$-Y-$CuO_2$ | 96.36 |
| $YBa_2Cu_3O_{6.60}$ | 63 | 2.2324 | 8.6271 | BaO-CuO-BaO | $CuO_2$-Y-$CuO_2$ | 64.77 |
| $LaBa_2Cu_3O_{7-\delta}$ | 97 | 2.1952 | 5.7983 | BaO-CuO-BaO | $CuO_2$-La-$CuO_2$ | 98.00 |
| $YBa_2Cu_4O_8$ (12 GPa) | 104 | 2.1658 | 5.5815 | BaO-CuO-CuO-BaO | $CuO_2$-Y-$CuO_2$ | 103.19 |
| $Tl_2Ba_2CuO_6$ | 80 | 1.9291 | 8.0965 | BaO-TlO-TlO-BaO | $CuO_2$ | 79.86 |
| $Tl_2Ba_2CaCu_2O_8$ | 110 | 2.0139 | 5.7088 | BaO-TlO-TlO-BaO | $CuO_2$-Ca-$CuO_2$ | 108.50 |
| $Tl_2Ba_2Ca_2Cu_3O_{10}$ | 130 | 2.0559 | 4.6555 | BaO-TlO-TlO-BaO | $CuO_2$-Ca-$CuO_2$-Ca-$CuO_2$ | 130.33 |
| $TlBa_2CaCu_2O_{7-\delta}$ | 103 | 2.0815 | 5.7111 | BaO-TlO-BaO | $CuO_2$-Ca-$CuO_2$ | 104.93 |
| $TlBa_2Ca_2Cu_3O_{9+\delta}$ | 133.5 | 2.0315 | 4.6467 | BaO-TlO-BaO | $CuO_2$-Ca-$CuO_2$-Ca-$CuO_2$ | 132.14 |
| $HgBa_2Ca_2Cu_3O_{8+\delta}$ | 135 | 1.9959 | 4.6525 | BaO-$HgO_x$-BaO | $CuO_2$-Ca-$CuO_2$-Ca-$CuO_2$ | 134.33 |
| $HgBa_2Ca_2Cu_3O_{8+\delta}$ (25 GPa) | 145 | 1.9326 | 4.4664 | BaO-$HgO_x$-BaO | $CuO_2$-Ca-$CuO_2$-Ca-$CuO_2$ | 144.51 |
| $HgBa_2CuO_{4.15}$ | 95 | 1.9214 | 7.0445 | BaO-$HgO_x$-BaO | $CuO_2$ | 92.16 |
| $HgBa_2CaCu_2O_{6.22}$ | 127 | 2.039 | 4.8616 | BaO-$HgO_x$-BaO | $CuO_2$-Ca-$CuO_2$ | 125.84 |
| $La_{1.837}Sr_{0.163}CuO_{4-\delta}$ | 38 | 1.7828 | 18.6734 | La/SrO-La/SrO | $CuO_2$ | 37.47 |
| $La_{1.8}Sr_{0.2}CaCu_2O_{6\pm\delta}$ | 58 | 1.7829 | 11.9900 | La/SrO-La/SrO | $CuO_2$-Ca-$CuO_2$ | 58.35 |
| $(Sr_{0.9}La_{0.1})CuO_2$ | 43 | 1.7051 | 17.6668 | Sr/La | $CuO_2$ | 41.41 |
| $Ba_2YRu_{0.9}Cu_{0.1}O_6$ | 35 | 2.0809 | 18.6123 | BaO | ½ $(YRu_{0.9}Cu_{0.1}O_4)$ | 32.21 |
| $(Pb_{0.5}Cu_{0.5})Sr_2(Y,Ca)Cu_2O_{7-\delta}$ | 67 | 1.9967 | 9.2329 | SrO-Pb/CuO-SrO | $CuO_2$-Y/Ca-$CuO_2$ | 67.66 |
| $Bi_2Sr_2CaCu_2O_{8+\delta}$ (unannealed) | 89 | 1.795 | 8.0204 | SrO-BiO-BiO-SrO | $CuO_2$-Ca-$CuO_2$ | 86.65 |
| $(Bi,Pb)_2Sr_2Ca_2Cu_3O_{10+\delta}$ | 112 | 1.6872 | 6.5414 | SrO-BiO-BiO-SrO | $CuO_2$-Ca-$CuO_2$-Ca-$CuO_2$ | 113.02 |
| $Pb_2Sr_2(Y,Ca)Cu_3O_8$ | 75 | 2.028 | 8.0147 | SrO-PbO-Cu-PbO-SrO | $CuO_2$-Y/Ca-$CuO_2$ | 76.74 |
| $Bi_2(Sr_{1.6}La_{0.4})CuO_{6+\delta}$ | 34 | 1.488 | 24.0797 | SrO-BiO-BiO-SrO | $CuO_2$ | 34.81 |
| $RuSr_2GdCu_2O_8$ | 50 | 2.182 | 11.3699 | SrO-$RuO_2$-SrO | $CuO_2$-Gd-$CuO_2$ | 50.28 |
| $La(O_{0.92-y}F_{0.08})FeAs$ | 26 | 1.7677 | 28.4271 | ½ (As-2Fe-As) | ½ (La-2O/F-La) | 24.82 |
| $Ce(O_{0.84-y}F_{0.16})FeAs$ | 35 | 1.6819 | 19.9235 | ½ (As-2Fe-As) | ½ (Ce-2O/F-Ce) | 37.23 |
| $Tb(O_{0.80-y}F_{0.20})FeAs$ | 45 | 1.5822 | 17.2624 | ½ (As-2Fe-As) | ½ (Tb-2O/F-Tb) | 45.67 |
| $Sm(O_{0.65-y}F_{0.35})FeAs$ | 55 | 1.667 | 13.2895 | ½ (As-2Fe-As) | ½ (Sm-2O/F-Sm) | 56.31 |
| $(Sm_{0.7}Th_{0.3})OFeAs$ | 51.5 | 1.671 | 14.3711 | ½ (As-2Fe-As) | ½ (Sm/Th-2O-Sm/Th) | 51.94 |
| $(Ba_{0.6}K_{0.4})Fe_2As_2$ | 37 | 1.932 | 17.4816 | As-2Fe-As | Ba/K | 36.93 |
| $Ba(Fe_{1.84}Co_{0.16})As_2$ | 22 | 1.892 | 28.0043 | As-2(Fe/Co)-As | Ba | 23.54 |
| $FeSe_{0.977}$ (7.5 GPa) | 36.5 | 1.424 | 23.8828 | $Se_{0.997}$ | Fe | 36.68 |
| $Fe_{1.03}Se_{0.57}Te_{0.43}$ (2.3 GPa) | 23.3 [a] | 1.597 | 30.4467 | $Se_{0.57}$-$Fe_{0.03}$-$Te_{0.43}$ | $Fe_{1.0}$ | 25.65 |
| $K_{0.83}Fe_{1.66}Se_2$ | 29.5 | 2.0241 | 20.4923 | Se-$Fe_{1.66}$-Se | $K_{0.83}$ | 30.07 |
| $Rb_{0.83}Fe_{1.70}Se_2$ | 31.5 | 2.1463 | 18.2889 | Se-$Fe_{1.70}$-Se | $Rb_{0.83}$ | 31.78 |
| $Cs_{0.83}Fe_{1.71}Se_2$ | 28.5 | 2.3298 | 18.1873 | Se-$Fe_{1.71}$-Se | $Cs_{0.83}$ | 29.44 |
| κ–[BEDT-TTF]$_2$Cu[N(CN)$_2$]Br | 10.5 | 2.4579 | 43.7194 | S-chains [BEDT-TTF]$_2$ | Cu[N(CN)$_2$]Br | 11.61 |

[a] $T_C$ = 23.3 K given in [26]; $T_C$ = 26.2 K for $Fe_{\sim 1}Se_{0.5}Te_{0.5}$ is given in [30].

For Fe:Se stoichiometry near 1:1 (x ~ 0, y ~ 0), $Fe_{1+x}Se_{1-y}$ has been found to exhibit defect-induced secondary magnetic phases [11] and weak ferromagnetism [12], as well as pressure-dependent antiferromagnetic fluctuations [2]. Principle features of the alkali-intercalated iron chalcogenide superconductors are coexisting antiferromagnetism with Néel temperatures well above room temperature [13], observations of nano-scale phase separation [14], indications of intrinsic percolative superconductivity [15], ordered Fe-vacancy microstructures [16], and relatively high normal-state resistivity just above $T_C$ [$\rho(T_C^+) > 10^{-3}$ Ω cm] [8]. The disorder evident in the 11 and 122 Fe chalcogenides is substantially greater than in their Fe-pnictide counterparts, and the distinctively different magnetic and band properties of the Fe chalcogenides, when compared to Fe pnictides (see, e.g. [17] [18]), is consistent with a pairing mechanism that is not particularly sensitive to multi-band interactions or spin fluctuations, but rather is rooted in the layered charge structure common to all families of high-$T_C$ compounds.

Previously, it was shown that the transition temperature of high-$T_C$ superconductors depends on structural distances, ionic valences, and Coulomb coupling between electronic bands in adjacent, spatially separated layers [19]. Analysis of thirty-one high-$T_C$ materials in five structural and chemical family types (cuprates, ruthenates, rutheno-cuprates, iron pnictides, organics) has revealed that the optimal transition temperature $T_{C0}$ is given by the universal algebraic expression $T_{C0} = k_B^{-1}\beta/\ell\zeta$. Here, $\ell$ is the in-plane spacing between interacting charges within the layers, $\zeta$ is the transverse distance between interacting layers and $\beta$ is a universal constant that may be written as $\beta = e^2\Lambda$, where $\Lambda$ is equal to about twice the reduced electron Compton wavelength ($k_B$ is Boltzmann's constant and $e$ is the elementary charge). The relevant electronic and structural parameters for these compounds are given in table 1 (adapted from [19] and augmented herein to include five Fe-Se-based compounds). Non-optimum compounds in which sample degradation is evident typically exhibit $T_C$ below $T_{C0}$.

In section 2 we give a brief background of the theory given in [19], focusing the discussion on the issues pertaining to the Fe chalcogenides. Calculations of $\ell$, $\zeta$ and $T_{C0}$ for $Fe_{1+x}Se_{1-y}$ $Fe_{1+x}Se_{1-y}Te_y$ and $A_zFe_{2-x}Se_2$ (for A = K, Rb, or Cs) are given in section 3. We discuss the nature of the charge reservoirs and elemental superconducting structure as well as some materials issues regarding these compounds in section 4, and our conclusions are summarized in section 5.

## 2. Theoretical background

In a recent theoretical paper [19] advocating our interlayer Coulombic pairing mechanism for high-$T_C$ superconductivity, a unique relationship between $T_{C0}$, the crystal structure and ionic valences was found to be common among superconducting families considered. In this model, the high-$T_C$ electronic structure is divided into two reservoirs, type I and type II, populated with equal two-dimensional (2D) densities of opposite charge and of differing mass (only opposing charges have thus far been considered, but the theory does not preclude like charges). The inclusion of the 1111 (e.g. Ln(O/F)FeAs) and 122 iron pnictides strongly suggests that the binary and ternary Fe chalcogenides may also be high-$T_C$ in nature, mediated by Coulomb interactions. To prove this, it is first necessary to define the two reservoirs for each case and identify the interacting layers. For $A_zFe_{2-x}Se_2$ this identification follows directly from the 122 Fe pnictides [19] in which the type I reservoir structure is defined as As-2Fe-As and the type II reservoir structure is the layer containing the alkaline earth and alkali metals (e.g. Ba/K). The Coulombic interaction thus occurs between the nearest-neighbor Ba/K and As layers. In the case of the corresponding 122 chalcogenides, the type I reservoir structure is Se-$Fe_{2-x}$-Se, the type II reservoir structure is the alkali metal $A_z$ and the pairing interaction occurs between nearest neighbor $A_z$ and Se layers. This assignment also allows the negation of interactions between nearest-neighbor $Fe_{2-x}$ and Se layers which, due to the strong antiferromagnetism of the $Fe_{2-x}$ layers [20], would suppress pairing.

For $Fe_{1+x}Se_{1-y}$ and $Fe_{1+x}Se_{1-y}Te_y$, we note that the Fe layers are only weakly ferromagnetic [12], and therefore identify the type I reservoir structures as the $Se_{1-y}$ or $Se_{1-y}Te_y$ layers and associate the Fe layers with type II. The pairing interaction therefore occurs between adjacent type I and type II layers. These definitions are illustrated in figure 1 for both $Fe_{1+x}Se_{1-y}$ and $A_zFe_{2-x}Se_2$, which also shows the pairing interaction distance $\zeta$ and periodicity $d$ in each case. The type I reservoir structure in $Fe_{1+x}Se_{1-y}$ (figure 1(a)) is a single Se layer (denoted by the count $\nu = 1$); in $A_zFe_{2-x}Se_2$ (figure 1(b)) the type I reservoir structure comprises a central layer of $Fe_{2-x}$ that is bounded by two Se layers (counted as $\nu = 2$); and in $Fe_{1+x}Se_{1-y}Te_y$ it is the Se and Te layers enveloping an excess Fe dopant (not illustrated, details in section 3.2), counted as $\nu = 2$. The dimension $d$ shown in figure 1 measures the transverse periodicity corresponding to one formula unit and represents the minimum thickness containing complete type I and type II charge reservoirs.

From [19] the optimal transition temperature $T_{C0}$, corresponding to the highest $T_C$ for a given compound structure and doping, is given by the expression,

$$T_{C0} = k_B^{-1} e^2 \Lambda \,/\, \ell\zeta = k_B^{-1}\, \beta\, (\sigma\eta/A)^{1/2} / \,\zeta\,, \qquad (1)$$

where $\sigma$ is the fractional charge per formula unit in a layer of the type I reservoir (an outer layer in cases of $\nu = 2$), corresponding to basal plane area $A$. The interaction occurs between two charge reservoirs (type I and type II) of opposite sign, separated by the nearest-neighbor interaction distance $\zeta$, where $\eta$ is the number of charge-carrying layers in the type II reservoir, $\ell^{-2} \equiv \sigma\eta/A$ is the areal charge density per type I layer per formula unit for participating carriers, and $\beta$ (= 0.1075 ± 0.0003 eV Å$^2$) $\equiv e^2\Lambda$ is a universal constant. The optimization of the superconducting state is achieved when the interacting carrier densities associated with the type I and type II reservoirs are in equilibrium (see equation (2.3) of [19]), and superconductivity in these systems is possible for $\zeta/\ell \leq 1$ (equation (2.1) of [19]). Equation (1) was validated, and $\beta$ determined, by 31 high-$T_C$ compounds from five different superconducting families (cuprates, ruthenates, rutheno-cuprates, iron pnictides, organics) with $T_{C0}$ values ranging from 10 K to 150 K. For all the Fe-Se chalcogenides considered herein the type II reservoir structures comprise a single atomic monolayer ($\eta = 1$), e.g. Fe in $Fe_{1+x}Se_{1-y}$ and $Fe_{1+x}Se_{1-y}Te_y$, and A in $A_zFe_{2-x}Se_2$ (see figure 1).

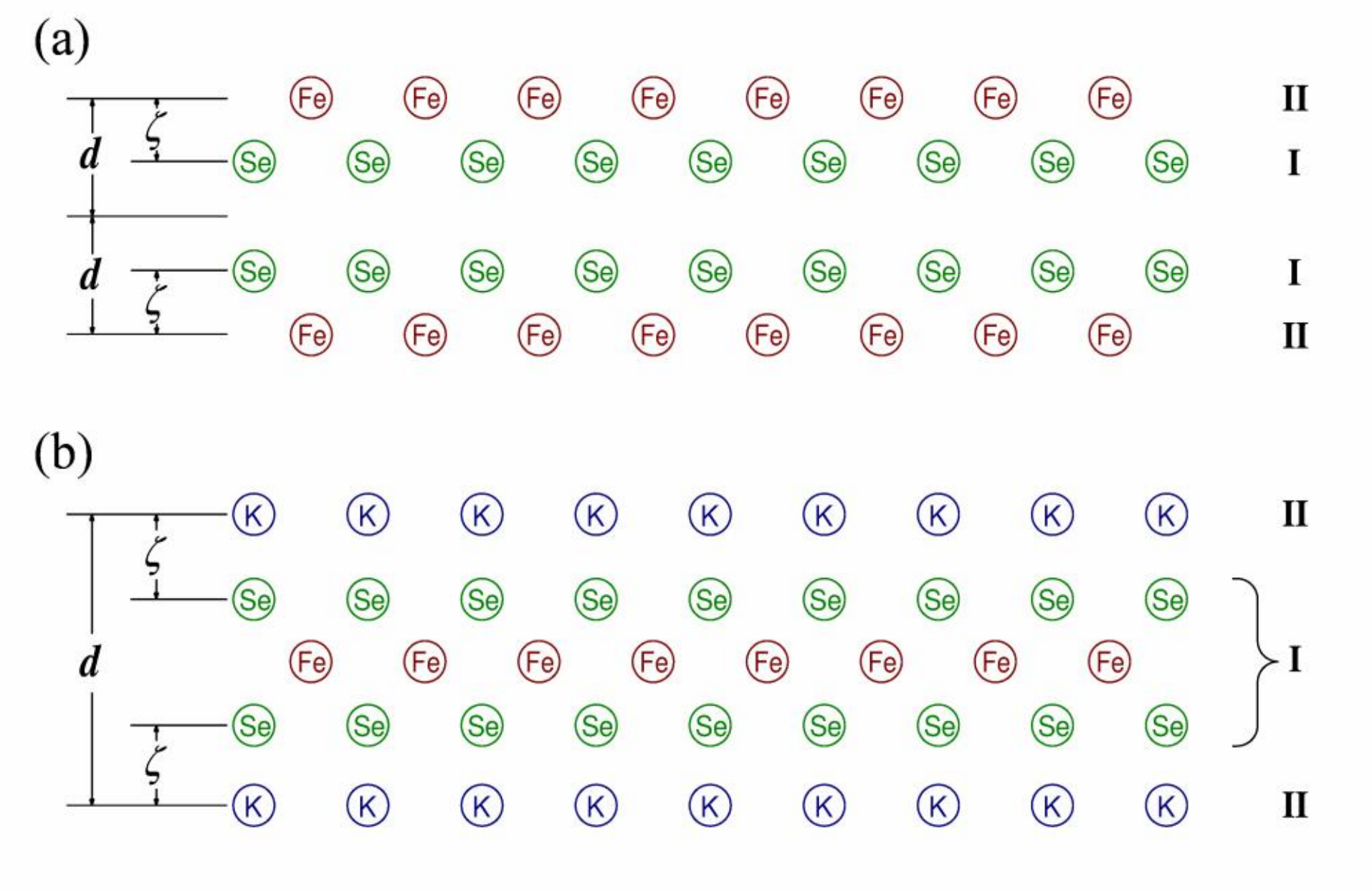

**Figure 1.** Schematic diagram of (a) $Fe_{1+x}Se_{1-y}$ and (b) $K_xFe_{2-y}Se_2$, as projected views along their [110] directions (linear densities of tetrahedrally-coordinated Fe are twice those of Se), illustrating the structures of the type I and type II reservoirs, the periodicity $d$ and the interaction distance $\zeta$. Vacancies and excess atoms are not shown.

Whereas $A$ and $\zeta$ are readily determined from crystal structure and $\eta$ is determined by inspection (here $\eta = 1$), the fractional charge $\sigma$ is much more elusive. In general, doping may be either cation or anion, and can occur in the type I reservoir, as in the case of $La_{2-x}Sr_xCuO_4$, the type II reservoir, e.g. $Ba_2Y(Ru_{1-x}Cu_x)O_6$ [19] or in both; the binary and ternary iron-chalcogenide systems (e.g. $A_xFe_{2-y}Se_2$) have as many as two dopant ions, one from each reservoir. Thus, for the compounds discussed herein, $\sigma$ can be determined by considering the cation and anion doping according to,

$$\sigma = \gamma \, [\, |v_I(x - x_0)_I| + |v_{II}(x - x_0)_{II}| \,], \qquad (2)$$

where $v_i$ is the net charge due to dopants (typically valence difference between dopant and the native ion) in reservoir $i$, $(x - x_0)_i$ is the generic doping factor in which $x$ denotes the content of the dopant species (e.g. as x in $A_xB$) and $x_0$ is the minimum value of $x$ required for superconductivity. Equation (2) is a generalization of equation (2.4$a$) from [19], which incorporates the possibility of non-unity valence doping and contains two terms corresponding to the two charge reservoirs; the absolute values confine the summation to the magnitudes of the individual contributions to $\sigma$. The factor $\gamma$ derives from the allocation of the dopant charge by considering a given compound's structure.

To calculate $\sigma$ for a given compound, the modifying factor $\gamma$ is determined by application of the set of rules developed in [19]; in treating doping according to equation (2), the following (first set) of charge allocation rules are used:

(1$a$) Sharing between N (typically 2) ions or structural layers introduces a factor of 1/N in $\gamma$.

(1$b$) The doping is shared equally between the hole and electron reservoirs, resulting in a factor of 1/2.

Multiple charge sources in opposing reservoirs (e.g. determined by A and Fe stoichiometries in $A_zFe_{2-y}Se_2$) are treated by equation (2) as a single contribution to $\sigma$ (hence the absolute values), with $\gamma$ determined by rules 1$a$ and 1$b$.

Numerous high-$T_C$ materials share the same value of $\sigma$ as optimal $YBa_2Cu_3O_{7-\delta}$, which we have denoted $\sigma_0$ and determined to have the value 0.228 for $\delta = 0.08$. For those compounds where the $(x - x_0)_i$ cannot be discerned independently through doping, $\sigma$ can be calculated by scaling to $\sigma_0$ for $YBa_2Cu_3O_{6.92}$ according to $\sigma = \gamma\, \sigma_0$, where $\gamma$ is defined in conjunction with a (second set) of valence scaling rules that are discussed in [19].

The two-dimensional carrier density is determined by the total areal charge associated with the type I reservoir (equivalently, the type II reservoir, as per rule *1b*), and is expressed as $n_{2D} = \nu\sigma/A$ [19]. The corresponding (zero-temperature) superfluid volume density, given by $n_{3D} = n_{2D}/d$, has been validated in [19] by comparison with magnetic penetration depth data for several cuprate, organic, and iron-pnictide superconductors. Absence of a scaling relationship between $T_{C0}$ and carrier densities (2D or 3D) was also noted in [19]; this is discussed for the case of the iron-based superconductors in section 4.

## 3. Determining $T_{C0}$ of the iron chacolgenide systems

The 11 and 122 Fe-Se chalcogenide compounds are very similar to 1111 and 122 Fe-As pnictides [19], and thus should follow similar behaviours with regard to charge allocation. Below we calculate $\gamma$, $\sigma$ and $T_{C0}$ for $Fe_{1+x}Se_{1-y}$, $Fe_{1+x}Se_{1-y}Te_y$ and $A_zFe_{2-x}Se_2$ (for A = K, Rb, or Cs), adding the results to table 1.

### *3.1 Tetragonal (11) $Fe_{1+x}Se_{1-y}$*

While superconducting (tetragonal) $Fe_{1+x}Se_{1-y}$ is very nearly stoichiometric [11] [21],[5] some degree of symmetry breaking, arising from either a Se deficiency or an excess in Fe content, must exist to induce the superconducting condensate. It is also clear that the optimal superconducting state is achieved at a hydrostatically applied pressure of 7.5-8.5 GPa [2] [3] [4], which we attribute to pressure-induced charge redistribution, analogous to the response of $YBa_2Cu_4O_8$ (see e.g. [22]). Although assigning reservoir types to the Fe and Se layers is somewhat subjective, it is also not a necessary requirement in determining σ (for convention, we refer to the negative valence Se layer as type I, as shown in figure 1(a)). Given the +2 and −2 valences of Fe and Se, respectively, the charge doping is 2x or 2y (or, typically, the sum thereof), where x and y are small and positive quantities.

Work presented in [3] follows the behavior of superconducting samples of $FeSe_{1-y}$ with nominal stoichiometry $FeSe_{0.977}$ and a tetragonal to hexagonal phase ratio of 77:23 (at ambient pressure), as a function of hydrostatic pressure. Since this stoichiometry lies in the mid-range of the ambient pressure limits determined in [11] and [21], this material represents an ideal test case. Resistivity just above $T_C$ measured at ambient pressure shows $\rho(T_C^+)$ = ~100 μΩ-cm [3], which compares well with the high-$T_C$ cuprates and is indicative of a high-quality material. At ambient pressure (0 GPa), $T_{C,onset} \approx 12$ K, with a resistive zero at $T_{C,zero} \approx 8$ K. With the application of hydrostatic pressure $T_C$ is observed to increase to a maximum (optimum) value of between 36 and 37 K at ~7.5 GPa, mirrored by a concomitant minimum in the relative transition width $\Delta T_C/T_C$, and both reflective of a near-optimal compound.

High resistivity (~33 mΩ cm at T ~ 9 K) observed in [23] confirms that stoichiometric FeSe is dominantly a non-metallic material (conductivity is 0.3% of $FeSe_{0.977}$ studied in [3]). One may therefore assume that the FeSe binary compound of precisely 1:1 stoichiometry is effectively an insulating material and that doping is achieved by introducing excess Fe or depleted Se. From rule *1b* the dopant charge populates both charge reservoirs, yielding $\gamma = 1/2$. For $FeSe_{0.977}$ ($y = 0.023$, $y_0 = 0$), the doping factor of equation (2) has the single term $[2(y - y_0)] = 2(0.023)$ and the fractional charge is then calculated to be $\sigma = 2(0.023)/2 = 0.023$. From figure 6 of [3] the basal-plane lattice parameter at 7.5 GPa is $a$ = 3.622 Å, corresponding to $A$ = 13.1189 Å$^2$; the interaction distance $\zeta$ = 1.424 Å (perpendicular distance measured between nearest-neighbor Fe and Se ions) at the same pressure is taken from figure 4(c) of [4]. Given these parameters, equation (1) evaluates to $T_{C0}$ = 36.68 K, which agrees well with the 36 – 37 K claimed by [3] and the 37 K cited by [4] for a presumably similar sample (unfortunately, the precise stoichiometry is not given).

The authors of [4] point to the large *c*-axis compressibility and the corresponding decrease in the interlayer distances with increasing pressure as being responsible for the increase in $T_C$. However, the measured changes in $\zeta$ (from about 1.45 to 1.422 Å for variations in pressure between 0 and 8 GPa) are far too small to explain the increase in $T_{C,onset}$ from 12 to 37 K (likewise for changes in the basal-plane area $A$). This leaves pressure-induced charge transfer as the operational mechanism driving this compound to its optimum superconducting state. This interpretation is supported by the decreasing $\Delta T_C/T_C$ as the pressure approaches 7.5 GPa from below.

### *3.2 The $Fe_{1+x}Se_{1-y}Te_y$ system*

While the isovalent substitution of $Te^{-2}$ for $Se^{-2}$ cannot introduce carriers into this alloy, the substantial difference in ionic radii (2.21 Å and 1.98 Å, respectively [5]) can produce significant stress within the crystalline structure; Tegel et al. [5] correctly point out that Se and Te, which occupy the 2*c* orthorhombic site, differ in heights (*z* co-ordinates) by as much as 0.24 Å at 0 GPa. Occupancy of Se and Te lattice positions is discussed in detail by Louca et al. [6]. Much like applying hydrostatic pressure, internal strain can aid in charge transfer (an interesting perspective of this effect for the cuprates is given by Brown [24]), which may explain the larger $T_C$ at 0 GPa (~14 K), when compared to $FeSe_{1-y}$ (8 K). From these considerations, the introduction of new carriers must arise from Fe non-stoichiometry [25], however slight. Unlike vacancies, the excess ions can exist anywhere within the unit cell and produce the doping factor 2x. In this case, the excess $Fe^{+2}$ ions are found to occupy the 2*c* site [26] [27]. Whereas [26] and [27] do not differentiate between the Se and Te *z* co-ordinates, the measurements of [5] place the excess iron at a height between that of the Se and Te. In the region of stoichiometry 0.3 < y < 0.9 the ternary $Fe_{1+x}Se_{1-y}Te_y$ compounds are a structurally distinct superconducting phase (i.e. effectively 111 compounds), when compared to the FeSe-like phase for 0 < y < 0.1 (formulations with 0.1 < y < 0.3 yield coexistence of the two phases) [28] [29].

[5] In [11] and [21] the maximum diamagnetic screening occurs for x = 0, y = 0.01 ± 0.02 (or equivalently x = 0.01± 0.02, y = 0) with an ambient pressure $T_{C,zero}$ of about 8 K.

To calculate $\sigma$ for this system (y > 0.3), one must consider the type I reservoir structure to be $Te_y$-$Fe_x$-$Se_{1-y}$, where $\nu = 2$ and the charge doping, supplied by the excess $Fe^{+2}$ ions, is shared equally between the $Te_y$ and $Se_{1-y}$ layers, with comparable occupancies since y ~ 1/2. The result for $\gamma$ is a factor of 1/2 (rule *1a*) multiplied by an additional factor of 1/2 associated with rule *1b*, yielding $\gamma = 1/4$. Therefore, $\sigma = 2x/4$.

Since accurate information on Fe stoichiometry is essential, we consider results reported for a sample with composition $Fe_{1.03}Se_{0.57}Te_{0.43}$, where x ≈ 0.03 was obtained from pressure- and temperature-dependent Rietveld refinements of synchrotron x-ray powder diffraction data [26]. From the above discussion, $\sigma = 2(0.03)/4 = 0.015$. The sample under study exhibited superconducting transitions (determined from magnetization onsets) of 13.9 K at ambient pressure, reaching a maximum $T_C$ = 23.3 K under hydrostatic pressure of 2.3 GPa. Unfortunately, only resistance and zero-field-cooled magnetization data are given, limiting our ability to quantify the sample's quality. In another work on a $FeSe_{0.5}Te_{0.5}$ sample (in this case Fe stoichiometry of unity was apparently presumed, leaving $\sigma$ undetermined), the superconducting transitions (onset values) varied from 13.5 K at ambient pressure, reaching a maximum $T_C$ of 25 – 26 K (26.2 K reported) at 2 GPa [30]. The sample in this case was judged by the authors to be of high quality, based on a 12% Meissner fraction (the field-cooled magnetization is about 38% of the zero-field-cooled magnetization; normal-state resistivity is 1.4 mΩcm extrapolated to $T_C$ at 2 GPa). Since one does not expect the structural parameters to differ greatly between these two samples, it would normally be reasonable to accept 25.5 K as representative of the highest $T_C$ attained for this compound. Given that the Fe content is unknown for the latter, results from [26] are used in table 1, with the understanding that the quoted transition temperature may be lower than optimum. To calculate $T_{C0}$ for the $Fe_{1.03}Se_{0.57}Te_{0.43}$ compound, we use data taken at 2.0 GPa [26]; $a$ = 3.7317 Å and $b$ = 3.7262 Å, giving a basal plane area (per formula unit) of $A$ = 13.9051 Å$^2$. We further estimate an average interaction distance $\langle\zeta\rangle$ = 1.597 Å (assuming [26] measures an average between $\zeta_{Se}$ and $\zeta_{Te}$). This value is corroborated by figure 3(c) of [31] which indicates $\zeta_{Te}$ = 1.7638 Å and $\zeta_{Se}$ = 1.4162 Å for a sample under pressure (presumably 2.0 GPa), corresponding to an average value $\langle\zeta\rangle$ = 1.59 Å. Using $\langle\zeta\rangle$ takes into account disorder in site occupancy and interactions contributed by the further Te ion, given that $(\zeta_{Te} - \zeta_{Se}) << \langle\zeta\rangle$. Evaluating equation (1) for the stoichiometry of the $Fe_{1.03}Se_{0.57}Te_{0.43}$ sample then gives an optimal $T_{C0}$ = 25.65 K, which agrees well with the range of reported $T_C$ values. Above ~2.5 GPa [26], the compound undergoes an orthorhombic to monoclinic transition, resulting in small abrupt changes to some of the relevant lattice parameters.

Although a trend where $T_C$ (optimized for pressure) decreases with Te content is observed [28], our results for $FeSe_{0.977}$ and $Fe_{1.03}Se_{0.57}Te_{0.43}$, both under hydrostatic pressure, show it to be completely explained by structural increases in $A$ and $\zeta$, both of which are y-dependent, and the change in $\nu$ from 1 in the binary compound to 2 in the ternary alloy.

*3.3 The (122) $A_zFe_{2-x}Se_2$ (A = K, Rb, or Cs) compounds*

Unlike $Fe_{1+x}Se_{1-y}$, the ternary $A_zFe_{2-x}Se_2$ series of compounds can be optimized at ambient pressure [32] [33],[6] and are characterized by a rather large normal-state resistivity at $T_C$, signifying a comparatively higher scattering rate. For these materials, the type I reservoir structures are identified as the Se-$Fe_{2-x}$-Se triple-layers, with ν = 2, and the alkali metal $A_z$ layers are defined as the type II (η = 1) reservoir structures [the exact formula-unit structure that we consider is: $Fe_{1-x/2}$-Se-$A_z$-Se-$Fe_{1-x/2}$, or equivalently, $A_{z/2}$-Se-$Fe_{2-x}$-Se-$A_{z/2}$, the latter corresponding to figure 1(b)]. There are generally two doping sources; one associated with the type I reservoir and the other with the type II. The two terms in equation (2) are $[z_0 - z]$ and $+2[x_0 - x]$ for $A_z$ and $Fe_{2-x}$, respectively, where the prefactor of 2 corresponds to the double valency of Fe (the terms $v_i(x - x_0)_i$ for the two reservoir types have opposite signs). There also appears to be (at least) two insulating end materials corresponding to the alkali metal and iron components (determined for A = K): $KFe_2Se_2$ [8] such that $z_0 = 1$, and $KFe_{2-x_0}Se_2$, where $0.40 \leq x_0 \leq 0.42$ [34] [35] [36][7]. Since x and z depend on growth stoichiometries, these two doping sources generally provide unequal contributions to the fractional charge, which is found by combining the magnitudes of the two charges as per equation (2), and treating doping as if from a single source. Thus, by analogy with $Ba_{1-x}K_xFe_2As_2$, the charge allocation rules *1a* and *1b* apply. Here, one factor of 1/2 arises from rule 1*b* requiring that the doping charge be shared between the two reservoirs; the factor of (1/2)(1/2) arises from dividing the charge among the four individual layers of the two Fe-Se structures of the type I reservoir (rule 1*a* for N=4, or applied twice with N=2). From this we can write $\gamma = (1/2)(1/2)(1/2) = 0.125$ [19], and

$$\sigma = 0.125\,[(1 - z) + 2(0.40 - x)]\,. \qquad (3)$$

To test equation (3) and thus our theory for the 122 Fe-Se chalcogenides, we begin with the potassium-intercalated compound that was studied by a number of authors (first reported in [7]). Results in [7], where an average stoichiometry was obtained from chemical analysis of boule crystals, show a broadened transition (onset $T_C \approx 30$ K) and high $\rho(T_C^+) \sim 0.6$ mΩcm (estimated by us from resistance and sample dimensions), which indicate that the sample produced for this first study is non-optimum for superconductivity. A later study on a sample with $T_C = 29.5$ K of composition $K_{0.83(2)}Fe_{1.66(1)}Se_2$ [i.e. z = 0.83(2) and x = 0.34(1)], as obtained by structural refinement of neutron powder diffraction data taken at 300 K, appears to be closest to optimum, as judged from a reasonable superconducting fraction (22% by magnetometry) and comparatively low magnetic moment amplitudes of 2.55 – 2.57 $\mu_B$ per Fe ($\mu_B$ is the Bohr magneton) [37].[8] From equation (3) one calculates $\sigma = 0.125\,[0.17 + 2(0.06)] = 0.0363$; assuming I4/m symmetry, $A = (3.9043\ \text{Å})^2 = 15.2436\ \text{Å}^2$ and $\langle\zeta\rangle = 2.0241$ Å, yielding from equation (1) $T_{C0} = 30.07$ K in excellent agreement with experiment.

Another superconducting composition for which we have good (unambiguous) information is $Rb_{0.83(1)}Fe_{1.70(1)}Se_2$ (stoichiometry from structural refinement analysis), having $T_C = 31.5$ K [37], which according to equation (3) gives, $\sigma = 0.125\,[0.17 + 2(0.10)] = 0.0463$; setting $x_0 = 0.40$ is consistent with observations that $Rb_{0.89}Fe_{1.58}Se_2$ is insulating [35] [41]. With $A = (3.9353\ Å)^2 = 15.4866\ Å^2$, $\langle\zeta\rangle = 2.1463$ Å for this rubidium-based compound, equation (1) predicts $T_{C0} = 31.78$ K. Finally, we consider the Cs-based compound $Cs_{0.83(1)}Fe_{1.71(1)}Se_2$ [42], stated as having a $T_C$ of 28.5 K. Earlier works on the same single crystals give $T_{C,onset} = 29.6$ K from d.c. susceptibility and resistivity [43] and an extrapolated $T_C = 27.4$ K from a.c. susceptibility [44]; an onset $T_C \sim 30$ K was also reported for a single crystal from the same batch [33]. Assuming an I4/m space group symmetry, refinement analysis of synchrotron x-ray diffraction data at 300 K gives $A = (4.0177\ Å)^2 = 16.1419\ Å^2$ and $\langle\zeta\rangle = 2.3298$ Å. From equation (3), $\sigma = 0.125\ [0.17 + 2(0.11)] = 0.0488$, which corresponds to $T_{C0} = 29.44$ K. Thus excellent agreement with experiment is found for both compounds.

[6] A slight increase in $T_C$ (by a couple of degrees K) with applied (hydrostatic) pressure as low as 0.8 GPa has been observed in $Cs_zFe_{2-x}Se_2$ [32], possibly indicative of a non-optimal stoichiometry or lattice frustration. Another study of a sample with $T_C$(0 GPa) ≈ 30 K shows $T_C$ decreasing with increasing pressure [33].

[7] The estimated size (5.5% Fe) of the Fermi surface pocket centered at $M$ (π,0) for $K_{0.8}Fe_{1.7}Se_2$ determined by angle-resolved photoemission spectroscopy (ARPES) in [36] is equivalent to σ = 0.055 which, from equation (3), corresponds to $x_0 = 0.42$.

[8] Other studies of $K_zFe_{2-x}Se_2$ reported: a novel large magnetic moment [9]; vacancy ordering (superconductivity data not presented) [38]; crystals with weak Meissner effects were studied in [39] (also high $\rho(T_C^+)$), [40] and [8].

Note that for $(z_0 - z) = 2(x_0 - x)$, the available 2D densities of electrons and holes in opposing reservoirs are automatically the same. In this case, the charge is distributed equally between the layers of the type I reservoir according to rule *1a* (rule *1b* would not apply), and one has $\sigma = 0.25\ [2(0.40 - x)] = 0.25\ (1 - z)$, for which $z = 0.2 + 2x$. While the stoichiometries of the three compounds studied do not reflect such an exact equality (the K compound has too little Fe (relative to K), and the Rb and Cs materials have comparatively too much Fe), they are rather close. Taking the average of all three, $\langle 1 - z\rangle = 0.17$, $\langle 2(0.4 - x)\rangle = 0.18 \pm 0.05$, and the doping ratio (Fe/A) is $1.06 \pm 0.31$. As in the case of other high-$T_C$ compounds, we expect that precise equilibrium is attained via charge transfer as the carriers condense to form a superconducting state [19].

In the quaternary $(Tl_{1-z}A_z)Fe_{2-x}Se_2$ compounds (see e.g. [45]), the Fe doping is typically increased relative to the corresponding ternary compound to compensate for the charge lost due to the isovalent substitutional (Tl/A) pair. This fact lends strong support to the validity of equation (3). We should also point out that the additional compositional variability introduced by the addition of $Tl_{1-z}$ may make an accurate determination of the stoichiometry more difficult (i.e. more accurately than presently attained).

## 4. Discussion

Table 1 presents data for the 31 high-$T_C$ compounds examined in prior work [19] (including seven Fe-As pnictide superconductors), plus the five Fe-Se superconductors examined in section 3 of this work. Listed are the compounds (organized into eight groups based on shared structural and/or doping properties), the measured $T_{C0}$, the specific lengths $\ell$ and $\zeta$, the atomic constituents of the types I and II charge reservoirs, and the theoretical $T_{C0}$ calculated from equation (1). These data are plotted in figure 2 as measured $T_{C0}$ *vs.* the derived quantity $(\ell\zeta)^{-1}$ with the Fe-based compounds indicated by distinctive symbols. The diagonal line represents equation (1) with the previously obtained value $\beta/k_B = 1247.4 \pm 3.7$ K-Å$^2$ with uncertainty in predicted $T_{C0}$ of $\pm 1.37$ K (from fitting 30 compounds – $Ba_2YRu_{0.9}Cu_{0.1}O_6$ was omitted from the fit owing to the relatively large uncertainty in $T_{C0}$) [19]. By including the five Fe chalcogenide compounds discussed herein and fitting all 36 compounds in table 1, the fitted slope is $\beta/k_B = 1246.7 \pm 3.6$ K-Å$^2$, with an uncertainty of $\pm 1.34$ K. Moreover, fixing $\beta/k_B$ at the previous value, one also finds an uncertainty in predicted $T_{C0}$ of 1.34 K, negating the need to reevaluate β.

Figure 3 provides a comparison between the present Coulomb interaction theory, as applied to the twelve Fe-based compounds in table 1, and an alternative hypothesis [17]; figure 3(a) is an expanded-scale view of $T_{C0}$ as a function of $(\ell\zeta)^{-1}$ (standard deviation 1.2 K of points from the theoretical line, reproduced from figure 2); figure 3(b) shows $T_{C0}$ against the superfluid volume density, expressed theoretically as $n_{3D} = \nu\ell^{-2}d^{-1}$, which produces substantial scatter among the data points (the dashed line serves as a guide to the eye; a two-parameter linear fit, not shown, yields standard deviation 9.2 K). Similarly, $T_{C0}$ when plotted against $n_{2D}$ or $\ell^{-2}$ is found to produce substantial point scatter as well. It is evident that thus far an alternative theoretical hypothesis that treats carrier densities and omits the interaction distance $\zeta$ does not accurately predict $T_{C0}$.

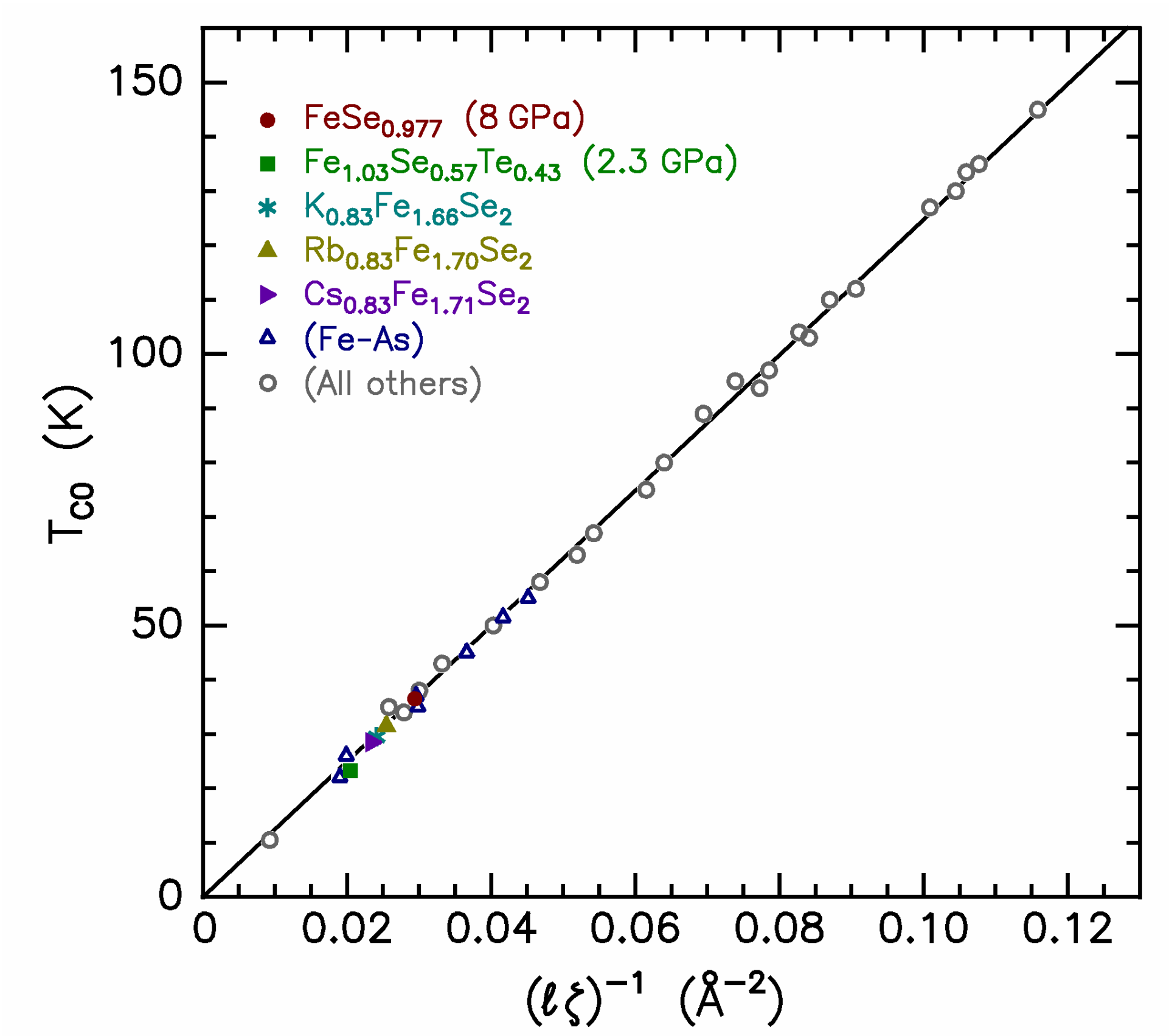

**Figure 2.** Optimal superconducting transition temperature $T_{C0}$ versus $(\ell\zeta)^{-1}$, where $\ell$ is the mean in-plane distance between participating charges and $\zeta$ is the distance between interacting layers. The five 11 and 122 Fe chalcogenides (solid symbols) are compared to the 1111 and 122 Fe pnictides (open triangles) and the remainder of the thirty-six compounds (open circles), exhibiting behaviours in agreement with equation (1), represented by solid line.

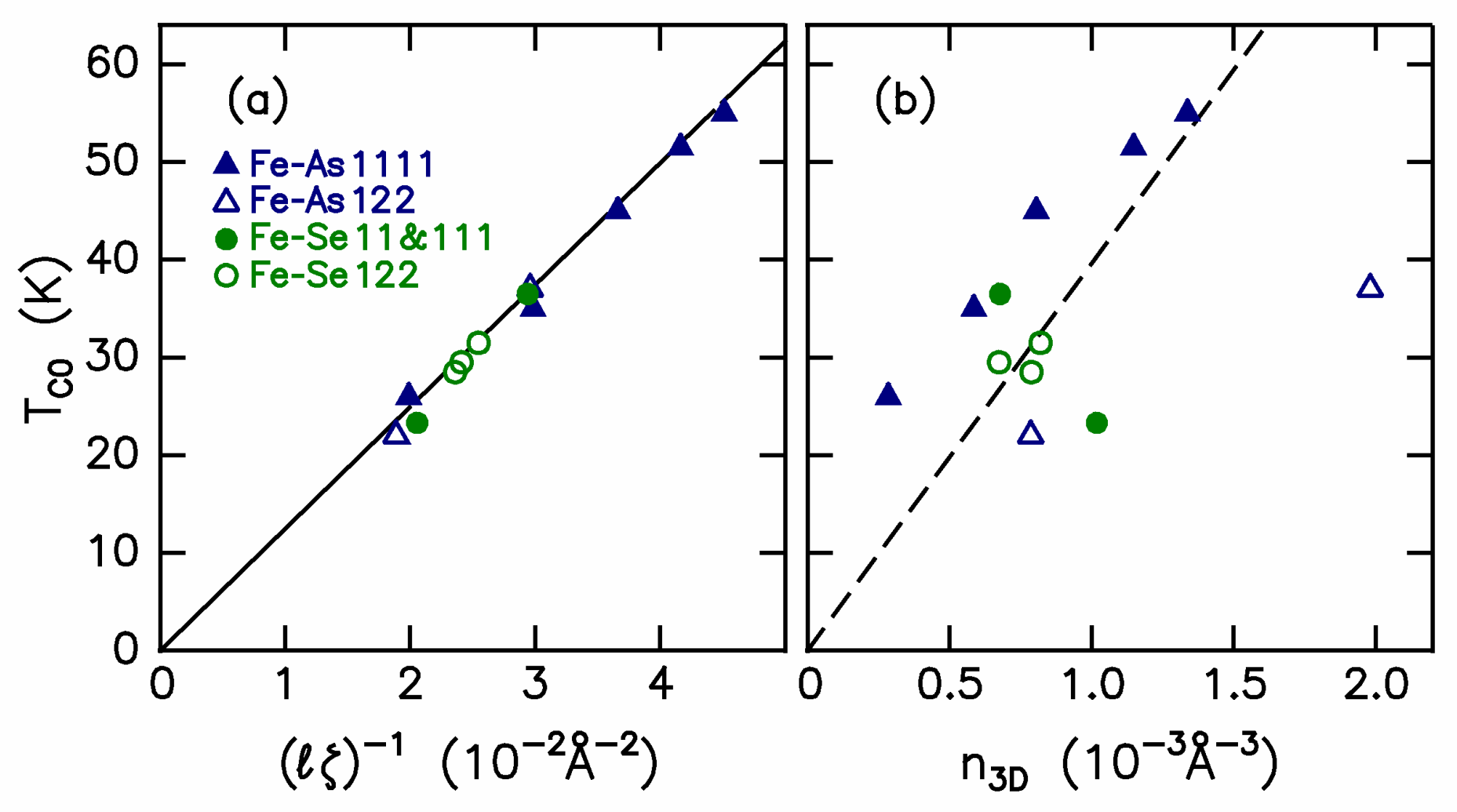

**Figure 3.** Comparative plots of $T_{C0}$ versus (a) $(\ell\zeta)^{-1}$ (line is theory) and (b) $n_{3D}$ (dashed line as guide to the eye) for the twelve Fe-based high-$T_C$ superconductors of table 1.

*4.1 Charge reservoir structure*

Identification of the layered charge reservoirs is key to accurately predicting transition temperatures, particularly in the case of the iron chalcogenides, which exhibit a greater richness in their superconducting structures. Although the iron-based superconductors contain the basic tetragonally co-ordinated As-2Fe-As or Se-2Fe-Se three-layer building block, they exhibit distinct differences in the structures of the charge reservoirs that yield high-$T_C$ superconductivity. In the 1111 and 122 pnictide high-$T_C$ compounds, which were treated previously [19], the type I reservoir structures were generally defined to be ½(As-2Fe-As) and As-2Fe-As, respectively, with the type II reservoir structure being the ½(Ln-2O/F-Ln) or Ba(K) (or similar) layers. The large (attractive) potential difference between $As^{-3}$ and the $Ln^{+3}$ ionic layers, and the relatively short intervening distance favor these two nearest-neighbor layer types for the locus of the pairing interaction.

For the $A_zFe_{2-x}Se_2$ chalcogenides, given their structural similarity to the 122 Fe pnictides, it becomes natural to define the type I reservoir structure as Se-$Fe_{2-x}$-Se and the alkali-metal A layer (with A = K, Rb or Cs) as the type II reservoir structure, with the pairing interaction occurring between the nearest-neighbor Se and A ionic layers. We note that this assignment is consistent with that of the cuprates, where the analogue of the BaO-CuO-BaO structure is Se-$Fe_{2-x}$-Se. More importantly, the large magnetic fields associated with the Fe cations (with local moments of 3.31 $\mu_B$/Fe [20] [46][9]), could act to suppress superconductivity if the Fe layers were directly involved in the pairing.

In binary $Fe_{1+x}Se_{1-y}$ the two charge reservoir structures are each monolayers, comprising Se or Fe for types I or II, respectively. In the ternary $Fe_{1+x}Se_{1-y}Te_y$, the type I reservoir structure is the $Se_{1-y}Te_y$ bilayer doped interstitially with $Fe_x$. In both $Fe_{1+x}Se_{1-y}$ and $Fe_{1+x}Se_{1-y}Te_y$, the Fe is only weakly ferromagnetic [12] [20]. Thus including the Fe layer as one of the two layers involved in the pairing interaction would introduce little or no pair-breaking perturbation.

*4.2 Elemental superconducting structure*

Given the high-$T_C$ mechanism as described previously in [19] and herein, the minimum number of layers required for superconductivity would incorporate a complete set of type I and type II reservoirs corresponding to the thickness *d* shown in figure 1. The two-dimensional nature of high-$T_C$ superconductivity has been validated in studies of ultra-thin layers fabricated with thickness approaching *d* (reviewed for the cuprates in [47], and in iron pnictide structures containing thick perovskite layers [48]). In the case of $YBa_2Cu_3O_{7-\delta}$, for example, these structures are contained within the thickness of the *c*-axis height of the unit cell ($d = c$), whereas for $Bi_2Sr_2CaCu_2O_{8+\delta}$ the minimum structure required is contained in one-half of a unit cell height ($d = c/2$).

For the Fe pnictides and chalcogenides, there are typically two formula units per unit cell such that the periodicity is one-half the *c*-axis height ($d = c/2$). Thus, for $Fe_{1+x}Se_{1-y}$ the minimal structure comprises one $Fe_{1+x}$ layer and one $Se_{1-y}$ layer; for $Fe_{1+x}Se_{1-y}Te_y$ the minimal structure is a single Fe layer and a Se/Te double layer with intervening Fe inclusions; and for $A_zFe_{2-x}Se_2$ the minimal structure required for superconductivity consists of a single alkali metal layer $A_z$ and an $Fe_{2-x}Se_2$ triple layer structure. The combination of these (type I and II) elemental superconducting structures is required for superconductivity, a fact that should be distinguished from a stand-alone Fe-anion tri-layer which forms the basis of several suggested pairing scenarios [49]. While the tri-layer As-2Fe-As structure is common to Fe-pnictide superconductors and the analogous Se-2Fe-Se structure is present in the binary Fe-Se and ternary Fe-Te-Se iron chalcogenides, the alkali-intercalated $A_zFe_{2-x}Se_2$ compounds stand as notable exceptions, owing to the Fe vacancies and their $\sqrt{5}\times\sqrt{5}$ superlattice ordering [16] [50] [41]. This has led to some speculation that different physics drives the superconductivity of the intercalated chalcogenides [50]. However, the obvious agreement with our interlayer Coulomb interaction model and correspondingly accurate predictions of $T_{C0}$ for all three chalcogenide structures (see figures 2 and 3(a)) renders such commentary rather moot.

[9] The Néel temperatures are found to be as high 559 K and the Fe magnetic moments form a collinear antiferromagnetic structure with the propagation wave vector oriented along ⟨101⟩ [46].

As noted in [19], of the two specific lengths that determine $T_{C0}$ according to equation (1), the interaction distance itself is comparatively weakly correlated with $T_{C0}$. For the Fe-based compounds in table 1, where ζ lies in the range 1.42 – 2.33Å, the linear correlation coefficient between $T_{C0}$ and $\zeta^{-1}$ is $R^2 = 0.14$. The stronger correlation is observed for ℓ, where for the Fe-based compounds ℓ lies in the range 13.29 – 30.45 Å and $R^2 = 0.79$. Additionally, the interaction distance ζ varies more broadly than the thickness of the As-2Fe-As or Fe-Te-Se building blocks, or equivalently, the height of the anion (e.g., As or Se) relative to neighboring Fe, as projected along the *z*-axis. As considered for about twenty Fe-based superconductors in [51], Fe-anion heights tend to cluster in the vicinity of 1.38 Å; the authors have sketched a peaked relation (no mathematical form given) with $T_C$ of width 0.12 Å (full width, half maximum) and suggested several trends. Owing to scatter in the data and the multivalued nature of the implied relations, the simple structural metric of Fe-anion height is insufficient for accurately predicting $T_C$. In earlier works limited to studies of iron pnictide compounds, $T_C$ was considered to be a simply monotonically increasing function of pnictogen-ion height [49] (opposite to the trends reported for chalcogen-ion heights in [51]), and for compounds with highest $T_C$ the Fe-pnictogen tetrahedron bond angle trends towards the regular value of 109.47 º [52,53]. Iron-anion heights turn out to be equivalent to ζ for only two cases, the iron chalcogenides $Fe_{1+x}Se_{1-y}$ and $Fe_{1+x}Se_{1-y}Te_y$. While the comparatively smaller $T_{C0}$ of $Fe_{1+x}Se_{1-y}Te_y$ does reflect the slightly larger ⟨ζ⟩, it is primarily due to a reduced σ resulting from the charge being distributed between both the $Se_{1-y}$ and $Te_y$ layers (i.e. due to a change from 11 to 111 structure, and not to a transition from tetragonal to monoclinic symmetry). Also, proposed correlation of $T_C$ with bond length or Se-Fe-Se bond angle, e.g. as noted by comparing the binary $Fe_{1+x}Se_{1-y}$ with the alloy in [6], are rather incidental when considering the distinctly different charge reservoir structures and doping mechanisms for these two compounds.

### *4.3 Percolation and phase separation*

Crystals of the intercalated ternary $A_zFe_{2-x}Se_2$ family of compounds grown to date indicate a coexistence of superconductivity and antiferromagnetism, as well as high normal-state electrical resistivities at temperatures just above $T_C$ (e.g. for A = K, Rb, Cs, (Tl,K), and (Tl,Rb), $\rho(T_C^+)$ = 6.1, 6.9, 41, 29, and 5.0 mΩ cm, respectively) [13]. Since a two-dimensional model of transport ought to be applicable, the normal-state resistivity may be expressed in the form $\rho(T_C^+) = (h/e^2)d/(k_F\ell_e)$, with h being Planck's constant, *d* the periodicity (see figure 1), and $(k_F\ell_e)$ the product of Fermi wavevector (in-plane, circular contour) and electron mean free path. Consider as an example the A = K compound, where from ARPES data, $k_F \approx 0.56a^{-1}$ has been estimated (see footnote 7) [36]; from structural data, $a$ = 3.9 Å and $d = c/2$ = 7.1 Å [9] we estimate a non-metallic $(k_F\ell_e) \approx 0.3$ for a $K_{0.8}Fe_{2-x}Se_2$ sample with $\rho(T_C^+)$ = 6.1 mΩ cm in [13] and a marginally metallic $(k_F\ell_e) = 1.2$ for one of the samples ($K_{0.82}Fe_{1.63}Se_2$ with $\rho(T_C^+)$ = 1.55 mΩ cm) in [7]. Similar analyses applied to the $Fe_{1+x}Se_{1-y}$ system find a non-metallic $(k_F\ell_e) \approx 0.02$ for the stoichiometric FeSe material in studied in [23] and a good-metal value $(k_F\ell_e) \approx 7$ for the superconducting compound $FeSe_{0.977}$ of [3].

In the case of $K_zFe_{2-x}Se_2$, the superconductivity apparently occurs dispersed within non-superconducting material. Evidence for inhomogeneous superconductivity is found in magnetization measurements; field-cooled diamagnetic moments of about 1% of zero-field-cooled moments are signatures of weak Meissner effects [9] [39] [40] [8] and low-field anomalies in hysteresis loops have been interpreted as evidence of intrinsic percolative superconductivity [15]. Intrinsic separation of magnetic and non-magnetic phases has been reported for similar crystals probed by focused (300 nm) x-ray diffraction [14]. Inhomogeneity presents some uncertainty in the optimum z and x values pertinent to the determination of charge fractions and calculation of theoretical $T_{C0}$ for the $A_zFe_{2-x}Se_2$ compounds.

As in all high-$T_C$ compounds, the superconducting condensate of the chalcogenides is dilute, requiring relatively few carriers as compared to conventional (phonon-mediated) superconductors. In the case of the ternary $A_zFe_{2-x}Se_2$ materials, the carriers available for superconductivity are introduced via non-stoichiometric deficiencies or excesses of ions; knowing accurately the optimum superconducting and end (insulating) materials compositions is essential to determining $T_{C0}$.[10] There are certain other Fe-chalcogenide superconductors, such as sulphur-substituted iron tellurides, where doping, e.g. by air exposure or ageing (reviewed in [54]), is thus far insufficiently characterized for quantitative analysis.

## 5. Conclusions

The superconducting transition temperatures $T_{C0}$ for five Fe-Se iron-chalcogenide compounds, namely $Fe_{1+x}Se_{1-y}$ and $Fe_{1+x}Se_{1-y}Te_y$, both optimized under pressure, and three 122 $A_zFe_{2-x}Se_2$ materials containing alkali intercalants A = K, Rb, or Cs, were considered from the perspective of our interlayer Coulombic coupling theory of high-$T_C$ superconductivity. Although these iron-containing superconductors display remarkable variety in their crystalline (see e.g. figure 1), magnetic and electronic band structures, they are shown to share the same physical mechanism for high-$T_C$ superconductivity as found for the Fe-As iron pnictides, cuprates, ruthenates, rutheno-cuprates, and organics considered previously [19]. Carriers and interacting charges in these compounds, quantified by the charge fraction $\sigma$ per type I interacting layer, are introduced by small stoichiometric differences that give rise to the superconducting states. In the case of $Fe_{1+x}Se_{1-y}$ the doping charge is produced by a small Fe excess (interstitials) as 2x or Se deficiency (vacancies) as 2y (valencies ±2, respectively); for the ternary alloy $Fe_{1+x}Se_{1-y}Te_y$, which has a structure distinctly different than $Fe_{1+x}Se_{1-y}$, the doping charge 2x is introduced as Fe interstitials within the $Se_{1-y}Te_y$ bilayer structure; and for the ternaries $A_zFe_{2-x}Se_2$, the doping charge is the sum $(z_0-z) + 2(x_0-x)$, where $z_0 = 1$ and $x_0 = 0.40$ are determined by insulating parent compounds. The excellent agreement between calculated and observed $T_{C0}$ (figures 2 and 3(a)) would tend to negate pairing schemes sensitive to spin-fluctuations [2] or multiband properties [49]. Moreover, the observation that $T_{C0}$ is essentially uncorrelated with $n_{3D}$ (figure 3(b)) effectively negates any hypothetical modeling of a dependence on superfluid density (e.g. putative linearities [17]). Comparison of these newer materials with the Fe-based pnictides [17] [18], possessing distinctly different band structure and doping characteristics,[11] and exhibiting unique magnetic behavior, suggest a pairing mechanism rooted in the layered charge structure common to all families of high-$T_C$ compounds.

The charge fraction $\sigma$ (= $\ell^{-2}A$) was determined for each of the subject compounds by utilizing the charge allocation rules from [19] (rules 1*a* and 1*b* in section 2) applied to their type I and II charge reservoirs (figure 1). Applying equation (1), optimal transition temperatures $T_{C0}$ for a binary sample of nominal stoichiometry $FeSe_{0.977}$, the Te-alloyed ternary $Fe_{1.03}Se_{0.57}Te_{0.43}$, and three representative 122 samples, $K_{0.83(2)}Fe_{1.66(1)}Se_2$, $Rb_{0.83(1)}Fe_{1.7(1)}Se_2$ and $Cs_{0.83(1)}Fe_{1.71(1)}Se_2$ were calculated and found to be in excellent agreement with experimental values (table 1; figures 2 and 3(a)). Inclusion of these five new compounds (table 1) decreases the uncertainty in calculated $T_{C0}$ slightly (from 1.37 to 1.34 K), with statistically unchanged value of the empirical universal constant β (1246.7 ± 3.7 KÅ$^2$). Thus, the number of high-$T_C$ compounds confirming the validity of equation (1) has been extended to 36, and the number of superconducting families explained has been increased to six.

---

[10] A recent paper [50] has suggested that all superconducting $A_zFe_{2-x}Se_2$ compounds are members of the same $A_2Fe_4Se_5$ family which share the same Fe vacancy and magnetic structure. However, the basic premise of the 2-4-5 structure being superconducting is effectively refuted by others [35] [41].

[11] Doping of the 1111 iron pnictides is typically accomplished in the type II layers by substituting $F^{-1}$ for $O^{-2}$, introducing O vacancies or direct cation doping of the lanthanide sites. For the 122 pnictides, doping involves cation substitution of the type II layer or the type I Fe layers. For $Fe_{1+x}Se_{1-y}$ and its Te-substituted derivative, excess Fe or Se/Te vacancies are required to induce superconductivity, while the doped charge in the related 122 chalcogenides originates from deficiencies in both the type II alkali metal layers and type I Fe layers.

*Acknowledgements*

We are grateful for the support of the Physikon Research Corporation (Project No. PL-206) and the New Jersey Institute of Technology. We would also like to thank Prof. Trevor Tyson for raising the question of how the Fe-Se 11 and 122 compounds fit into our high-$T_C$ picture. Publication of this work is to appear in Journal of Physics: Condensed Matter.